\documentstyle[preprint,prl,aps,epsfig,floats]{revtex}
\tightenlines
\begin{document}
\preprint{
\vbox{\halign{&##\hfil    \cr
        & CMU-HEP00-03    \cr
        & UTPT-00-07      \cr
        & August 2000        \cr
        }}}

\title{NRQCD Analysis of Bottomonium Production at the Tevatron}

\author{Eric Braaten}
\address{Physics Department, Ohio State University, Columbus, OH 43210, USA}

\author{Sean Fleming}
\address{Physics Department, University of Toronto, Ontario, M5S 1A7, Canada}

\author{Adam K. Leibovich}
\address{Department of Physics, Carnegie Mellon University, 
	Pittsburgh, PA 15213, USA}

\maketitle
\begin{abstract}
Recent data from the CDF collaboration on the production of 
spin-triplet bottomonium states at the Tevatron $p \bar p$ collider
are analyzed within the NRQCD factorization formalism.  
The color-singlet matrix elements are determined 
from electromagnetic decays and from potential models.
The color-octet matrix elements are determined by fitting the CDF data 
on the cross sections for $\Upsilon(1S)$, $\Upsilon(2S)$, 
and $\Upsilon(3S)$ at large $p_T$
and the fractions of  $\Upsilon(1S)$ coming from
$\chi_b(1P)$ and $\chi_b(2P)$.  
We use the resulting matrix elements to predict the cross sections 
at the Tevatron for the spin-singlet states $\eta_b(nS)$ and $h_b(nP)$.  
We argue that $\eta_b(1S)$ should be observable in Run II
through the decay $\eta_b \to J/\psi + J/\psi$.
\end{abstract}

\vfill \eject

\narrowtext
\section{Introduction}

The NRQCD factorization formalism provides a systematic framework for 
analyzing the inclusive production of heavy quarkonium \cite{B-B-L}.
Long-distance effects involving the binding of a heavy 
quark-antiquark pair into quarkonium are factored into
parameters called NRQCD matrix elements.
These nonperturbative parameters are universal,
so values extracted from one high energy physics experiment
can be used to predict the production rate in others.  
The NRQCD matrix elements scale as definite powers of $v$, 
where $v$ is the typical relative velocity of the heavy quark.
The NRQCD factorization approach becomes phenomenologically useful
upon truncating the expansion in $v$ so as to reduce the 
independent NRQCD matrix elements to a manageable number.
The truncation is most reliable for the heaviest quarkonium states,
namely the $b \bar b$ system for which $v^2$ is roughly 1/10.

The most abundant source of data on bottomonium production
is the Tevatron $p \bar p$ collider.  In Run IA of the Tevatron,
the CDF collaboration was able to resolve the individual S-wave
bottomonium states $\Upsilon(1S)$, $\Upsilon(2S)$, and $\Upsilon(3S)$
and measure their production cross sections \cite{CDF-upsIa}.
An analysis of the CDF data within the NRQCD factorization formalism
was carried out by Cho and Leibovich \cite{Cho-Leibovich}.
The analysis is complicated by the production of P-wave 
bottomonium states that subsequently make transitions to S-wave states.
Cho and Leibovich found that the CDF data was insufficient 
to determine all the important NRQCD matrix elements and
they had to make educated guesses for some of them.

The CDF collaboration has recently analyzed 
the data on bottomonium production from Run IB at the Tevatron. 
In addition to much higher statistics on the cross sections for 
$\Upsilon(1S)$, $\Upsilon(2S)$, and $\Upsilon(3S)$  \cite{CDF-upsIb}, 
they also have results on the production of the P-wave states
$\chi_b(1P)$ and $\chi_b(2P)$ \cite{CDF-chib}.
The high quality of the new CDF data justifies an updated   
theoretical analysis, with careful attention to the experimental and
theoretical errors. 

In this paper, we present a quantitative analysis of the new CDF data 
on bottomonium production within the NRQCD factorization formalism.
The color-singlet NRQCD matrix elements for S-wave states are determined 
from their electromagnetic decays, while those for P-wave states 
are estimated from potential models.
The color-octet NRQCD matrix elements are determined by fitting the CDF data,
taking full account of the feeddown from transitions between 
bottomonium states.  The resulting values of the matrix elements
are used to predict the cross sections for the spin-triplet
and spin-singlet bottomonium states in Run II of the Tevatron.

\section{NRQCD Matrix Elements}

The NRQCD factorization approach provides a model-independent framework
for analyzing the inclusive production of heavy quarkonium \cite{B-B-L}.
The factorization formula for the differential cross section for 
the inclusive production of a bottomonium state $H$ of momentum $P$
has the schematic form
\begin{equation}
d \sigma[H(P)] \;=\;
\sum_n d \sigma[b \bar b(n,P)] \langle O^H(n) \rangle,
\label{sig-NRQCD}
\end{equation}
where the sum extends over both color-singlet and color-octet 
and over all angular momentum channels for the $b \bar b$ pair.
The $b \bar b$ cross sections, which are independent of the 
bottomonium state $H$, can be calculated using perturbative QCD.
All dependence on the state $H$ is factored into parameters
$\langle O^H(n) \rangle$ called {\it NRQCD matrix elements}.
These phenomenological parameters can be expressed as matrix elements 
in an effective field theory called nonrelativistic QCD (NRQCD).
A nonperturbative analysis of NRQCD reveals
how the various matrix elements scale with the typical relative 
velocity $v$ of the heavy quark in quarkonium.  
Spin symmetry, which is an approximate symmetry of QCD,
also gives relations between various matrix elements.

The relative importance of the terms in the factorization formula
(\ref{sig-NRQCD}) depends on the size of the $b \bar b$ cross sections
and on the size of the matrix elements.
According to the velocity-scaling rules, 
the most important matrix element for direct
$\Upsilon(1S)$ production is the color-singlet parameter
$\langle O^{\Upsilon(1S)}_1(^3S_1) \rangle$.
The spin-symmetry relations can be used to reduce the next most 
important matrix elements to three color-octet parameters:
$\langle O^{\Upsilon(1S)}_8(^3S_1) \rangle$,
$\langle O^{\Upsilon(1S)}_8(^1S_0) \rangle$, and
$\langle O^{\Upsilon(1S)}_8(^3P_0) \rangle$.
These color-octet matrix elements are important, because the cross sections 
for producing color-octet $b \bar b$ pairs can be much larger than 
for color-singlet $b \bar b$ pairs.
There are analogous matrix elements that describe the direct production
of $\Upsilon(2S)$ and $\Upsilon(3S)$.
The NRQCD factorization formula (\ref{sig-NRQCD}) 
for direct $\Upsilon(nS)$ production reduces to
\begin{eqnarray}
d \sigma[\Upsilon(nS)] &=&
d \sigma[b \bar b_1(^3S_1)] \langle O^{\Upsilon(nS)}_1(^3S_1) \rangle
+ d \sigma[b \bar b_8(^3S_1)] \langle O^{\Upsilon(nS)}_8(^3S_1) \rangle
\nonumber
\\ 
&& + d \sigma[b \bar b_8(^1S_0)] \langle O^{\Upsilon(nS)}_8(^1S_0) \rangle
+ \left( \sum_J (2J+1) d \sigma[b \bar b_8(^3P_J)] \right)
	\langle O^{\Upsilon(nS)}_8(^3P_0) \rangle.
\label{sig-ups}
\end{eqnarray}
The factor of $2J+1$ in the last term comes from 
using a spin-symmetry relation to eliminate
$\langle O^{\Upsilon(nS)}_8(^3P_J) \rangle$ in favor of 
$\langle O^{\Upsilon(nS)}_8(^3P_0) \rangle$.

The most important matrix elements for the direct production of the 
P-wave states $\chi_{bJ}(1P)$, $J=0,1,2$,
can be reduced to a color-singlet parameter
$\langle O^{\chi_{b0}(1P)}_1(^3P_0) \rangle$ 
and a single color-octet parameter 
$\langle O^{\chi_{b0}(1P)}_8(^3S_1) \rangle$.
There are analogous matrix elements that describe the direct production
of $\chi_{bJ}(2P)$ and $\chi_{bJ}(3P)$.
The NRQCD factorization formula (\ref{sig-NRQCD}) 
for direct $\chi_{bJ}(nP)$ production reduces to
\begin{eqnarray}
d \sigma[\chi_{bJ}(nP)] &=&
d \sigma[b \bar b_1(^3P_J)] 
	\langle O^{\chi_{bJ}(nP)}_1(^3P_J) \rangle
+ (2J+1) d \sigma[b \bar b_8(^3S_1)] 
	 \langle O^{\chi_{b0}(nP)}_8(^3S_1) \rangle.
\label{sig-chi}
\end{eqnarray}
In the last term, the factor of $2J+1$ comes from using a
spin-symmetry relation to eliminate
$\langle O^{\chi_{bJ}(nP)}_8(^3S_1) \rangle$ in favor of
$\langle O^{\chi_{b0}(nP)}_8(^3S_1) \rangle$.
We can also use a spin-symmetry relation to replace
$\langle O^{\chi_{bJ}(nP)}_1(^3P_J) \rangle$ in the first term by 
$(2J+1) \langle O^{\chi_{b0}(nP)}_1(^3P_0) \rangle$.
The matrix elements for $\Upsilon(nS)$ and $\chi_{bJ}(nP)$ 
enumerated above should be sufficient for a quantitative description
of the production of S-wave and P-wave bottomonium states.

The NRQCD factorization formula gives the cross section for the 
{\it direct} production of a given bottomonium state.  
The cross sections that are most easily measured in experiments are 
{\it inclusive} cross sections that include contributions from the 
direct production of higher bottomonium 
states which subsequently decay into the given state.
For example, the feeddown from
$\chi_b(1P)$, $\Upsilon(2S)$, and $\chi_b(2P)$ accounts for roughly
27\%, 11\%, and 11\% of the $\Upsilon(1S)$ cross section, 
respectively \cite{CDF-chib}.  Taking into account the feeddown from
higher $\Upsilon(mS)$ and $\chi_{bJ}(mP)$ states,
the cross section for inclusive $\Upsilon(nS)$ production 
can be written 
\begin{eqnarray}
d \sigma[\Upsilon(nS)]_{\rm inc} &=&
d \sigma[b \bar b_1(^3S_1)] 
	\langle O_1(^3S_1) \rangle^{\Upsilon(nS)}_{\rm inc}
+ \sum_J d \sigma[b \bar b_1(^3P_J)] 
	\langle O_1(^3P_J) \rangle^{\Upsilon(nS)}_{\rm inc}
\nonumber
\\ 
&& + d \sigma[b \bar b_8(^3S_1)] 
	\langle O_8(^3S_1) \rangle^{\Upsilon(nS)}_{\rm inc}
+ d \sigma[b \bar b_8(^1S_0)] 
	\langle O_8(^1S_0) \rangle^{\Upsilon(nS)}_{\rm inc}
\nonumber
\\
&& 
+ \left( \sum_J (2J+1) d \sigma[b \bar b_8(^3P_J)] \right)
	\langle O_8(^3P_0) \rangle^{\Upsilon(nS)}_{\rm inc},
\label{sig-total}
\end{eqnarray}
where the ``inclusive NRQCD matrix elements" are
\begin{equation}
\langle O[n] \rangle^{\Upsilon(nS)}_{\rm inc} \;=\;
\sum_H B_{H \to \Upsilon(nS)} \langle O^H[n] \rangle.
\label{O-total}
\end{equation}
The sum over $H$ includes $\Upsilon(nS)$ and all higher 
bottomonium states that can make transitions to $\Upsilon(nS)$.
The coefficient $B_{H \to H'}$ is the inclusive branching fraction 
for $H$ to decay into $H'$. 
By convention, we define $B_{H \to H} = 1$.
The inclusive branching fraction for the observed bottomonium states 
are collected in Table~\ref{tab:branch}.  
These numbers were obtained by combining 
the measured branching fractions for the exclusive decays
$\Upsilon(nS) \to \chi_{bJ}(mP)+ \gamma$,
$\chi_{bJ}(nP) \to \Upsilon(mS) + \gamma$, and 
$\Upsilon(nS) \to \Upsilon(mS) + \pi \pi$,
with the exception of $B_{\Upsilon(3S) \to \Upsilon(1S)}$,
which is a direct measurement \cite{PDG}.

\begin {table}
\begin {center}
\begin {tabular}{l|c|ccc|c|ccc|c}
& $\Upsilon(3S)$ & $\chi_{b2}(2P)$ & $\chi_{b1}(2P)$ & $\chi_{b0}(2P)$ &
  $\Upsilon(2S)$ & $\chi_{b2}(1P)$ & $\chi_{b1}(1P)$ & $\chi_{b0}(1P)$ &
  $\Upsilon(1S)$ \\
\hline
$\chi_{bJ}(3P)$ 
& 0 ? &   0 ?      &      0 ?     &     0 ?     &      0 ?     &
          0 ?      &      0 ?     &     0 ?     &      0 ?     \\ 
\hline
$\Upsilon(3S)$  
& 1 & 11.4$\pm$0.8 & 11.3$\pm$0.6 & 5.4$\pm$0.6 & 10.6$\pm$0.8 & 
       0.6$\pm$0.1 &  0.6$\pm$0.1 & 0.4$\pm$0.1 & 11.2$\pm$0.5 \\
\hline
$\chi_{b2}(2P)$ 
&   &       1      &              &             & 16.2$\pm$2.4 &
       1.1$\pm$0.2 &  1.1$\pm$0.2 & 0.7$\pm$0.2 & 12.1$\pm$1.3 \\ 
$\chi_{b1}(2P)$ 
&   &              &       1      &             &   21$\pm$4   &     
       1.4$\pm$0.3 &  1.4$\pm$0.3 & 0.9$\pm$0.3 & 15.0$\pm$1.8 \\ 
$\chi_{b0}(2P)$ 
&   &              &              &       1     &  4.6$\pm$2.1 &     
       0.3$\pm$0.1 &  0.3$\pm$0.1 & 0.2$\pm$0.1 &  2.3$\pm$0.9 \\ 
\hline
$\Upsilon(2S)$  
&   &              &              &             &       1      & 
       6.6$\pm$0.9 &  6.7$\pm$0.9 & 4.3$\pm$1.0 & 31.1$\pm$1.6 \\
\hline
$\chi_{b2}(1P)$
&   &              &              &             &              &  
            1      &              &             &   22$\pm$4   \\ 
$\chi_{b1}(1P)$ 
&   &              &              &             &              &     
                   &       1      &             &   35$\pm$8   \\ 
$\chi_{b0}(1P)$ 
&   &              &              &             &              &
                   &              &       1     &     $<$ 6    \\ 
\end {tabular}
\end {center}
\caption{ \label {tab:branch}
	Inclusive branching fractions $B_{H \to H'}$ (in \%)
	for transitions between spin-triplet bottomonium states.
	The entries ``0 ?'' in the first row indicate that
	the feeddown from $\chi_{bJ}(3P)$ is neglected in our analysis. }
\end {table}

In Table~\ref{tab:branch}, we have not included the spin-singlet states
$\eta_b(nS)$ and $h_b(nP)$, which have yet to be observed.  
Transitions between spin-singlet and spin-triplet states are suppressed,
because they proceed through magnetic $\Delta S = 1$ transitions.
The rates for $\Delta S = 1$ transitions 
are suppressed relative to those for $\Delta S = 0$ transitions
by a factor of $v^2$, which is roughly an order of magnitude.
The branching fractions for $\eta_b(2S)$ into other bottomonium states 
are further suppressed by its large annihilation width into two gluons.
Quantitative estimates of the electromagnetic and hadronic transition
rates are given in Refs. \cite{Kuang-Yan,Riska}.
They support the conclusion that the branching fractions for decays 
of spin-singlet states into spin-triplet states can be neglected.

The tiny branching fractions in Table~\ref{tab:branch}
for the transition $\chi_{bJ}(2P) \to \chi_{bJ'}(1P)$ 
are the contributions from the double radiative transitions via $\Upsilon(2S)$.
We have not included the contributions 
from the two-pion decays
$\chi_{bJ}(2P) \to \chi_{bJ'}(1P) + \pi \pi$, which have not been observed.
We can estimate their magnitude by observing that the rates for
$\Upsilon(3S) \to \Upsilon(2S) + \pi \pi$ and 
$\Upsilon(3S) \to \Upsilon(2S) + \gamma \gamma$
are equal to within experimental errors.  
Since the phase space available for the transitions
$\chi_{bJ}(2P) \to \chi_{bJ'}(1P)$ is similar to that for
$\Upsilon(3S) \to \Upsilon(2S)$,  we expect the rate for 
$\chi_{bJ}(2P) \to \chi_{bJ'}(1P) + \pi \pi$ to be comparable to that for
$\chi_{bJ}(2P) \to \chi_{bJ'}(1P) + \gamma \gamma$.
Including the effects of two-pion transitions could increase 
the branching fraction for $\chi_{bJ}(2P) \to \chi_{bJ'}(1P)$ 
by a factor of 2 or 3, but since the values of
$B_{\chi_{bJ}(2P) \to \chi_{bJ'}(1P)}$ in Table~\ref{tab:branch} are all 
less than 1.5\%, they should still be negligible.

As indicated by the entries ``0 ?'' in the first row of Table~\ref{tab:branch}, 
we neglect the feeddown from the $\chi_b(3P)$ states, 
which have not been observed.  A naive extrapolation from the other 
entries of the Table suggest that the branching fractions for
$\chi_{b1}(3P)$ and $\chi_{b2}(3P)$ into $\Upsilon(3S)$ could be about 
12\%, while their branching fractions into $\Upsilon(1S)$ could be 
about 7\%.  These are small enough that they would not have a significant 
effect on our analysis.  
We have also neglected the feeddown from D-wave states.

\section{Parton Differential Cross Sections}

In hadron collisions, bottomonium with transverse momentum 
$p_T$ of order $m_b$ or larger is produced, at leading order in $\alpha_s$, 
by parton {\it fusion} processes $i j \to b \bar b + k$.  
The differential cross section for producing a bottomonium state $H$
with momentum $P$ can be expressed 
in the schematic form
\begin{equation}
d \sigma[H(P)]_{\rm fusion} \;=\; 
f_{i/p} \otimes f_{j/\bar p} \otimes 
	d \hat \sigma[i j \to b \bar b(P,n)+k] 
	\langle O^H(n) \rangle,
\label{sig-fusion}
\end{equation}
where there is an implied sum over the partons $i,j,k$ 
and over the $b \bar b$ channels $n$.

The order-$\alpha_s^3$ fusion cross section in (\ref{sig-fusion}) 
gives a good first approximation only if the transverse momentum 
is not too much larger or too much smaller than $m_b$.
For $p_T \gg m_b$, the order-$\alpha_s^3$ fusion cross section 
for the channel $b \bar b_8(^3S_1)$
has the scaling behavior $d \hat \sigma/dp_T^2 \sim 1/p_T^4$,
while all other channels are suppressed by powers of $m_b^2/p_T^2$
at leading order.  Parton processes with scaling behavior 
are called {\it fragmentation} processes.
The fragmentation contributions to $b \bar b$ channels other than 
$b \bar b_8(^3S_1)$ enter at higher order in $\alpha_s$.
The order-$\alpha_s^3$ fusion cross sections therefore
underestimate the $b \bar b$ cross section in these channels
at large $p_T$.
However, the CDF data on bottomonium production extends only out to 
$p_T = 20$ GeV  \cite{CDF-upsIb}, which is not large enough 
for fragmentation effects to dominate.  In extracting the NRQCD matrix 
elements from that data, it should therefore be sufficient 
to use the fusion cross section (\ref{sig-fusion}). 

The order-$\alpha_s^3$ fusion cross section in (\ref{sig-fusion}) 
also fails at small $p_T$.  
For some $b \bar b$ channels, including 
$b \bar b_8(^1S_0)$ and $b \bar b_8(^3P_{0,2})$,
there is an order-$\alpha_s^2$ fusion cross section 
from the parton process $i j \to b \bar b$, which produces 
a $b \bar b$ pair with $p_T=0$.
In these channels, the order-$\alpha_s^3$ fusion cross sections 
$d \sigma/dp_T^2$ diverge like $1/p_T^2$ as $p_T \to 0$.  
The divergence in the integral of the cross section 
for $i j \to b \bar b + k$ is cancelled by the radiative corrections
to the cross section for $i j \to b \bar b$, 
so that the cross section integrated 
over $p_T$ is finite order by order in $\alpha_s$.
In order to obtain a smooth prediction for $d \sigma/dp_T^2$
in the small $p_T$ region, it is necessary to resum higher order 
corrections involving soft-gluon radiation.  
This resummation will have a significant effect on the shape 
of the $p_T$ distribution, and therefore on the values of the 
NRQCD matrix elements used to fit that distribution.
We will avoid the complications due to soft-gluon radiation at small $p_T$
by using only the data from $p_T > 8$ GeV to fit the NRQCD matrix elements. 

We proceed to describe each of the factors in the fusion cross section
(\ref{sig-fusion}) in more detail.  
We include the contributions from the following combinations of 
colliding partons: $ij$ = $gg$, $gq$, $g \bar q$,  
$q \bar q$, where $q = u,d,s,c$. 
The parton distributions
$f_{i/p}(x_1,\mu_F)$ and $f_{j/\bar p}(x_2,\mu_F)$, which
specify the momenta of the colliding partons, depend on a
factorization scale $\mu_F$.  
We will consider the CTEQ5L and MRST98LO parton distribution functions.
They are both obtained from leading order analyses, and thus can be used 
consistently with leading order parton cross sections.
Explicit expressions for the parton differential cross sections $d\hat \sigma$
are given in Ref. \cite{Cho-Leibovich} 
and in Ref. \cite{B-K-V}.
They are proportional to $\alpha_s^3(\mu_R)$, where $\mu_R$ is the 
renormalization scale, and they depend on the mass $m_b$ of the bottom quark.
As part of the theoretical error,
we will allow $\mu_F$ and $\mu_R$ to vary by factors of 2 from
the central values $\mu_T = (m_b^2 + p_T^2)^{1/2}$.
This central value interpolates between half the partonic invariant mass
at $b \bar b$ threshold and half the partonic invariant mass
at large $p_T$ and central rapidity.

The cross sections also depend on two fundamental QCD parameters:
$\alpha_s$ and $m_b$.  We take the QCD coupling 
constant $\alpha_s(\mu)$ to run according to the one-loop formula,
with the boundary value appropriate to the 
parton distribution function that is being used.
For CTEQ5L, the coupling constant satisfies 
$\alpha_s(M_Z) = 0.127$ and $\alpha_s(m_b) = 0.232$.
For MRST98LO, the coupling constant satisfies 
$\alpha_s(M_Z) = 0.125$ and $\alpha_s(m_b) = 0.226$.

The other fundamental QCD parameter in our calculation 
is the bottom quark mass $m_b$.
There have been several recent determinations of $m_b$
using sum rules calculated to next-to-next-to-leading order accuracy
with nonrelativistic resummation \cite{mb-pole}.
A useful summary of these results is given by Beneke in Ref. \cite{beneke}.
The value of the pole mass
is rather unstable under radiative corrections compared to short-distance 
definitions of the mass, such as the running mass evaluated at its own scale,
$\bar m_b = m_b(\bar m_b)$.
Beneke's best estimate for this mass is $\bar m_b = 4.23 \pm 0.08 $ GeV. 
The definition of $\bar m_b$ is purely 
mathematical in character and not related to any physical thresholds
involving the $b$ quark.  Two definitions that are also relatively stable 
under radiative corrections and whose definitions
are related to thresholds in the bottomonium system are the 
1S mass, which is the perturbative energy of the lowest bound state,
and the PS mass, which is the sum of the pole mass and some 
energy related to the potential between the $b$ and $\bar b$.
Beneke's best estimates for these masses are
$m_{b,1S} = 4.77 \pm 0.11 $ GeV and
$m_{b,PS}(2 \, {\rm GeV}) = 4.57 \pm 0.10 $ GeV.
The relation between the two is given by a power series in $\alpha_s$:
\begin{equation}
m_{b,PS}(\mu) \;=\; 
m_{b,1S} - {4 \alpha_s(\mu) \over 3 \pi} \mu + O(\alpha_s^2),
\label{mass}
\end{equation}
The difference between Beneke's values for $m_{b,1S}$ and
$m_{b,PS}$ is mostly accounted for by the order-$\alpha_s$ correction.
We will choose the 1S mass as our prescription for the $b$ quark mass.
Beneke's central value for the PS mass differs by 2 standard deviations
from the 1S mass.
This difference should not be regarded as an ambiguity in the quark mass,
because it could not be eliminated by a more precise determination 
of $m_b$.  Instead its effects on the cross section could be 
decreased by calculating the 
next-to-leading order radiative correction to the parton cross sections.
The uncertainty due to different prescriptions for the quark mass
can therefore be regarded as part of the error due to
radiative corrections.

\section{Color-singlet Matrix Elements}

The color-singlet matrix elements for $\Upsilon(nS)$ can 
be determined phenomenologically from its decay rate into
a lepton pair.
The electronic decay rate of the $\Upsilon(nS)$, including the QCD
radiative correction of order $\alpha_s$ 
and the first relativistic correction of order $v^2$, is
\begin{equation}
\Gamma[\Upsilon(nS) \to e^+ e^-] \;=\;
{2\pi\alpha^2 e_b^2 \over 9m_b^2} 
\left( 1 - {8 \over 3} {\alpha_s \over \pi} 
	- {1 \over 3} {M_{\Upsilon(nS)} - 2 m_b \over 2 m_b} \right)^2
\langle O^{\Upsilon(nS)}_1(^3S_1) \rangle,
\label{ups-ee}
\end{equation}
where $e_b = -1/3$ is the bottom quark charge
and $m_b$ is the 1S mass.
The relativistic correction was first expressed in terms of 
$M_{\Upsilon(nS)} - 2 m_b$ 
by Gremm and Kapustin \cite{Gremm-Kapustin}.
The vacuum saturation approximation,
which is accurate up to corrections of order $v^4$,
has been used to express the NRQCD matrix element 
that enters naturally in annihilation rates in terms of the 
corresponding production matrix element 
$\langle O^{\Upsilon(nS)}_1(^3S_1) \rangle$. 
It has also been used to express the radiative and relativistic 
correction factor as a square.
Setting $m_b=4.77$ GeV and $\alpha_s(m_b) = 0.22$ 
and using the measured value for the decay rates, we obtain the 
values for the color-singlet matrix elements in Table~\ref{tab:CS-direct}.
In addition to the experimental errors in the decay rates,
there are theoretical errors from relativistic corrections and from
perturbative corrections.
As a measure of the relativistic error of order $v^4$, we take
the square of the largest of the order-$v^2$ corrections for the 
three $\Upsilon(nS)$ states: 
$[(M_{\Upsilon(3S)} - 2 m_b)/(3 m_b)]^2 \approx 0.2$\%.
As a measure of the perturbative error from higher orders in $\alpha_s$,
we take the square of the order-$\alpha_s$ correction in (\ref{ups-ee}):
$[16 \alpha_s/(3 \pi)]^2 \approx 14$\%.
The error bars quoted in Table~\ref{tab:CS-direct} are obtained
by combining the experimental, relativistic, and perturbative errors in 
quadrature. 
The error bars are dominated by the 14\% perturbative error, 
except in the case of the $\Upsilon(3S)$ 
for which the experimental error is 16\%.
The values for $\langle O^{\Upsilon(nS)}_1(^3S_1) \rangle$
in Table~\ref{tab:CS-direct} are larger by a factor of 3 than those 
given in Table I of the first paper in Ref. \cite{Cho-Leibovich} 
because of a normalization error in the Table.
This normalization error did not appear in the cross sections
and therefore did not affect the results.

\begin {table}
\begin {center}
\begin {tabular}{l|c|c|c||l|c|c}
     & \multicolumn{3}{c||}{$\langle O^{\Upsilon(nS)}_1(^3S_1) \rangle$} & 
     & \multicolumn{2}{c}{$\langle  O^{\chi_{b0}(nP)}_1(^3P_0) \rangle$} \\
     & phenomenology & potential models & lattice 
     & & potential models & lattice \\
\hline
$\Upsilon(3S)$ &  4.3 $\pm$ 0.9  &  3.7 $\pm$ 1.5  & 9.6 $\pm$ ? & 
$\chi_b(3P)$   &  2.7 $\pm$ 0.7  &   \\
$\Upsilon(2S)$ &  4.5 $\pm$ 0.7  &  5.0 $\pm$ 1.8  & 3.6 $\pm$ ? & 
$\chi_b(2P)$   &  2.6 $\pm$ 0.5  &   \\
$\Upsilon(1S)$ & 10.9 $\pm$ 1.6  & 10.8 $\pm$ 5.5  & 7.6 $\pm$ ? & 
$\chi_b(1P)$   &  2.4 $\pm$ 0.4  & 1.5 $\pm$ ? \\
\end {tabular}
\end {center}
\caption{ \label {tab:CS-direct}
	Direct color-singlet matrix elements for 
	$\Upsilon$ and $\chi_b$ states 
	($\langle O_1(^3S_1) \rangle$ in units of GeV$^3$,
	 $\langle O_1(^3P_J) \rangle$ in units of GeV$^5$).}
\end {table}

There is no data that can be used for phenomenological determinations 
of the color-singlet matrix elements for the P-wave states.
However the color-singlet matrix elements for both the S-wave and 
P-wave states can be estimated using wavefunctions from potential models.
Using the vaccuum-saturation approximation,
the color-singlet matrix element for $\Upsilon(nS)$
can be expressed in terms of its radial wavefunction at the origin,
while that for $\chi_{bJ}(nP)$ can be expressed in terms 
of the derivative of its radial wavefunction at the origin:
\begin{eqnarray}
\langle O^{\Upsilon(nS)}_1(^3S_1) \rangle
&\approx& {9 \over 2 \pi} |R_{\Upsilon(nS)}(0)|^2,
\label{O1-ups}
\\
\langle O^{\chi_{bJ}(nP)}_1(^3P_J) \rangle
&\approx& (2 J + 1) {9 \over 2 \pi} |R_{\chi_b(nP)}'(0)|^2.
\label{O1-chi}
\end{eqnarray}
Eichten and Quigg have tabulated the radial wavefunctions 
and their derivatives at the origin for 4 potential models 
that reproduce the observed bottomonium spectrum \cite{Eichten-Quigg}.
As estimates of the color-singlet matrix elements (\ref{O1-ups})
and (\ref{O1-chi}), we take the mean values from the 4 potential 
models in Ref. \cite{Eichten-Quigg}.
The mean values are tabulated in Table~\ref{tab:CS-direct}. 
The errors are the root-mean-square 
deviations of the 4 potential-model values.
The potential model estimates of
$\langle O^{\Upsilon(nS)}_1(^3S_1) \rangle$
are consistent with the phenomenological values, but have larger error bars.
This gives us some confidence in the potential-model
estimates for $\langle O^{\chi_{b0}(nP)}_1(^3P_0) \rangle$
in Table~\ref{tab:CS-direct}.
These values are consistent within errors with those used
in the analysis of Ref. \cite{Cho-Leibovich}.

The most accurate determination of the color-singlet matrix elements for 
the lowest bottomonium states may eventually come from lattice gauge theory.
The corresponding annihilation matrix elements can be readily calculated 
using lattice simulations of NRQCD \cite{NRQCD}.
The NRQCD collaboration has calculated the wavefunctions for the lowest 
S-wave states \cite{NRQCD} and P-wave states \cite{private}
of bottomonium.  Inserting these wavefunctions into the expressions
in (\ref{O1-ups}) and (\ref{O1-chi}), 
we obtain the estimates of the color-singlet matrix elements
in Table~\ref{tab:CS-direct}.  
The largest errors in the lattice calculations
come from matching of lattice NRQCD operators 
with continuum NRQCD operators and from the omission of dynamical
light quarks.  Both errors could be as large as 25\% in the present 
calculations.
It is premature to quote error bars for the
lattice gauge theory results in Table~\ref{tab:CS-direct}.

We will adopt the phenomenological values of
$\langle O^{\Upsilon(nS)}_1(^3S_1) \rangle$ in Table~\ref{tab:CS-direct} 
and the potential-model values for 
$\langle O^{\chi_{b0}(nP)}_1(^3S_1) \rangle$.
Using the branching fractions in Table~\ref{tab:branch}, 
we can form the linear combinations that appear in the expressions 
(\ref{sig-total}) for the inclusive $\Upsilon(nS)$ cross sections. 
These are tabulated in Table~\ref{tab:CS-total}.
As indicated by the zeros in the first row, we neglect feeddown from the 
$\chi_b(3P)$ states.

\begin {table}
\begin {center}
\begin {tabular}{l|c|c|c|c}
& $\langle O_1(^3S_1) \rangle^{\Upsilon(nS)}_{\rm inc}$ 
& $\langle O_1(^3P_0) \rangle^{\Upsilon(nS)}_{\rm inc}$ 
& ${1 \over 3} \langle O_1(^3P_1) \rangle^{\Upsilon(nS)}_{\rm inc}$
& ${1 \over 5} \langle O_1(^3P_2) \rangle^{\Upsilon(nS)}_{\rm inc}$ \\
\hline
$\Upsilon(3S)$ &   4.3  $\pm$ 0.9 &  0 ?  &  0 ?  &  0 ?  \\
$\Upsilon(2S)$ &   5.0  $\pm$ 0.7   &  0.12 $\pm$ 0.06  & 
                   0.55 $\pm$ 0.15  &  0.42 $\pm$ 0.10  \\
$\Upsilon(1S)$ &  12.8  $\pm$ 1.6   &       $<$ 0.2     &  
                   1.23 $\pm$ 0.25  &  0.84 $\pm$ 0.15     \\
\end {tabular}
\end {center}
\caption{ \label {tab:CS-total}
	Inclusive color-singlet matrix elements for $\Upsilon$ states 
	($\langle O_1(^3S_1) \rangle$ in units of GeV$^3$,
	 $\langle O_1(^3P_J) \rangle$ in units of GeV$^5$).}
\end {table}

\section{Color-octet Matrix Elements}

The color-octet NRQCD matrix elements are phenomenological
parameters that can only be determined from experimental data. 
We first extract the inclusive color-octet matrix elements for $\Upsilon(nS)$ 
from the CDF measurements of the inclusive $\Upsilon(nS)$ cross sections.
We then extract direct color-octet matrix elements for $\chi_{bJ}(nP)$ 
from the CDF measurements of the fraction of $\Upsilon(1S)$
coming from $\chi_b$'s.  This gives us enough information to determine
the direct color-octet matrix elements for $\Upsilon(nS)$ .

\subsection{Inclusive Matrix Elements for S-waves}

The inclusive $\Upsilon(nS)$ cross sections depend linearly
on the inclusive matrix elements defined in (\ref{O-total}).
The inclusive color-singlet matrix elements are given in 
Table~\ref{tab:CS-total}. 
We can extract the inclusive color-octet matrix elements from the 
CDF measurements of the $\Upsilon(nS)$ cross sections \cite{CDF-upsIb}.
The differential cross sections integrated over rapidities $|y| < 0.4$
have been measured out to $p_T =$ 20 GeV \cite{CDF-upsIb}.
The CDF data on $B d \sigma/d p_T$ for $\Upsilon(nS)$, 
where $B$ is the branching fraction for $\Upsilon(nS) \to \mu^+ \mu^-$,
are shown in Figures 1, 2, and 3 for $\Upsilon(1S)$, 
$\Upsilon(2S)$, and $\Upsilon(3S)$, respectively.
We avoid the problem of carrying out soft-gluon resummation
to determine the shapes of the theoretical $p_T$ distributions 
at low $p_T$ by using
only the data from $p_T >$ 8 GeV 
to fit the color-octet matrix elements. 
This leaves 5 $p_T$ bins for $\Upsilon(1S)$ 
and 3 $p_T$ bins each for $\Upsilon(2S)$ and $\Upsilon(3S)$.

The inclusive $\Upsilon(nS)$ cross sections depend on the 
inclusive color-octet matrix elements through the linear combination 
$[\langle O_8(^1S_0)) + m \; \langle O_8(^3P_0) \rangle/m_b^2]
	+ n \; \langle O_8(^3S_1))$, 
where $m$ varies from 4.6 at $p_T = 8$ GeV 
to 3.4 at $p_T = 20$ GeV,
while $n$ varies from 1.0 at $p_T = 8$ GeV 
to 6.3 at $p_T = 20$ GeV.
The parameters $\langle O_8(^1S_0) \rangle$ and
$\langle O_8(^3P_0) \rangle$
can not be determined independently, because the corresponding 
parton cross sections have similar dependences on $p_T$.
We therefore carry out our analysis under the two extreme assumptions
that either the $\langle O_8(^1S_0) \rangle$ term or the
$\langle O_8(^3P_0) \rangle$ term dominates and that the other 
can be neglected.  
Assuming both matrix elements are positive, the truth will
be somewhere in between the two extremes.  
We will take the difference between the two extremes as part of the 
theoretical error. 

For a given choice of the parton distributions
and the scales $\mu_F$ and $\mu_R$,
we can integrate the $\Upsilon(nS)$ differential cross section 
(\ref{sig-total}) over $|y|<0.4$ and over each $p_T$ bin.  
We determine the best fits for the color-octet matrix elements 
by minimizing the $\chi^2$ associated with the sum over $p_T$ bins.
Fixing $\mu_F =\mu_R = \mu_T$ and assuming that $\langle O_8(^3P_0) \rangle$
is negligible, we obtain the values of
$\langle O_8(^3S_1) \rangle$ and $\langle O_8(^1S_0) \rangle$
in the 1st and 3rd columns of Table~\ref{tab:CO-total}.
Repeating the analysis but assuming that $\langle O_8(^1S_0) \rangle$
is negligible, we obtain the values of 
$\langle O_8(^3S_1) \rangle$ and $\langle O_8(^3P_0) \rangle$
in the 2nd and 4th columns.  The first errors in Table~\ref{tab:CO-total}
are extracted from the matrix of second derivatives of the $\chi^2$ function. 
There is also an error from varying the renormalization and 
factorization scales $\mu_F$ and $\mu_R$. 
These errors are large, but since they are highly
correlated, we have separated them out as a second error in  
Table~\ref{tab:CO-total}.
The upper and lower errors are the shifts in the matrix elements 
that minimize $\chi^2$ when $\mu_R$ and $\mu_F$ are changed from the 
central value $\mu_T = \sqrt{m_b^2 + p_T^2}$ by multiplicative factors 
of 2 and 1/2, respectively.
The error from varying $m_b$ is also highly correlated, but it is smaller 
and we do not list it separately.  It can be taken into account when we 
use the matrix elements to calculate other observables.

\begin {table}
\begin {center}
\begin {tabular}{r|c|c|c|c}
& \multicolumn{2}{c|}{CTEQ5L}
& \multicolumn{2}{c}{MRSTLO} \\
\hline
$\langle O_8(^3S_1) \rangle^{\Upsilon(3S)}_{\rm inc}$ 
& $ 3.6 \pm 1.9^{+1.8}_{-1.3} $ & $ 3.9 \pm 1.7^{+2.0}_{-1.4} $
& $ 3.7 \pm 2.1^{+1.7}_{-1.3} $ & $ 4.1 \pm 1.9^{+1.9}_{-1.4} $
\\
$\langle O_8(^1S_0) \rangle^{\Upsilon(3S)}_{\rm inc}$ 
& $ 5.4 \pm 4.3^{+3.1}_{-2.2} $ & $ 0 $  
& $ 7.5 \pm 4.9^{+3.4}_{-2.5} $ & $ 0 $\\
${5 \over m_b^2}\langle O_8(^3P_0) \rangle^{\Upsilon(3S)}_{\rm inc}$ 
& $ 0 $ & $ 5.7 \pm 4.6^{+3.3}_{-2.3} $
& $ 0 $ & $ 7.9 \pm 5.2^{+3.7}_{-2.6} $\\
\hline
$\langle O_8(^3S_1) \rangle^{\Upsilon(2S)}_{\rm inc}$ 
& $ 18.0 \pm 5.6^{+8.9}_{-6.4} $&$ 17.2 \pm 5.0^{+8.7}_{-6.2} $
& $ 19.6 \pm 6.3^{+8.9}_{-6.5} $&$ 19.0 \pm 5.6^{+8.7}_{-6.4} $
\\
$\langle O_8(^1S_0) \rangle^{\Upsilon(2S)}_{\rm inc}$ 
& $ -10.2 \pm 9.7^{-3.1}_{+1.8} $ & $ 0 $  
& $ -8.7 \pm 11.1^{-2.4}_{+1.8} $ & $ 0 $\\
${5 \over m_b^2}\langle O_8(^3P_0) \rangle^{\Upsilon(2S)}_{\rm inc}$ 
& $ 0 $ & $ -10.6 \pm 10.2^{-3.0}_{+2.2} $
& $ 0 $ & $ -8.9 \pm 11.7^{-2.5}_{+1.8} $\\
\hline
$\langle O_8(^3S_1) \rangle^{\Upsilon(1S)}_{\rm inc}$ 
& $ 11.6 \pm 2.7^{+5.9}_{-4.2} $&$ 12.4 \pm 2.5^{+6.6}_{-4.7} $
& $ 11.7 \pm 3.0^{+5.7}_{-4.2} $&$ 13.0 \pm 2.8^{+6.4}_{-4.7} $
\\
$\langle O_8(^1S_0) \rangle^{\Upsilon(1S)}_{\rm inc}$ 
& $ 10.9 \pm 6.2^{+10.2}_{-7.1} $ & $ 0 $ 
& $ 18.1 \pm 7.2^{+11.4}_{-8.1} $ & $ 0 $\\
${5 \over m_b^2}\langle O_8(^3P_0) \rangle^{\Upsilon(1S)}_{\rm inc}$ 
& $ 0 $ & $ 11.1 \pm 6.5^{+10.7}_{-7.5} $
& $ 0 $ & $ 18.6 \pm 7.5^{+11.9}_{-8.4} $\\
\end {tabular}
\end {center}
\caption{ \label {tab:CO-total}
	Inclusive color-octet matrix elements for $\Upsilon$ states
	(in units of $10^{-2}$ GeV$^3$).}
\end {table}

\begin{figure}[ht]
\centerline{\epsfxsize 11cm \epsffile{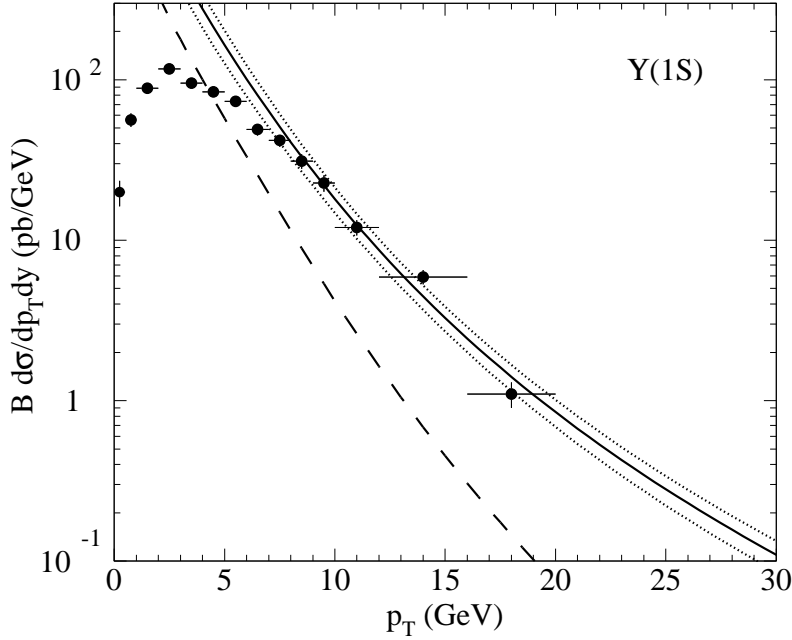}}
\caption[Inclusive cross section for $\Upsilon(1S)$ at Run I.]
{Inclusive cross section for $\Upsilon(1S)$ at $y = 0$ in Run I
	multiplied by its branching fraction $B$ into $\mu^+ \mu^-$
	as a function of $p_T$:  CDF data, 
	NRQCD fit (solid line) with statistical error bars (dotted lines),
	and color-singlet model prediction (dashed line).}
\end{figure}

\begin{figure}[ht]
\centerline{\epsfxsize 11cm \epsffile{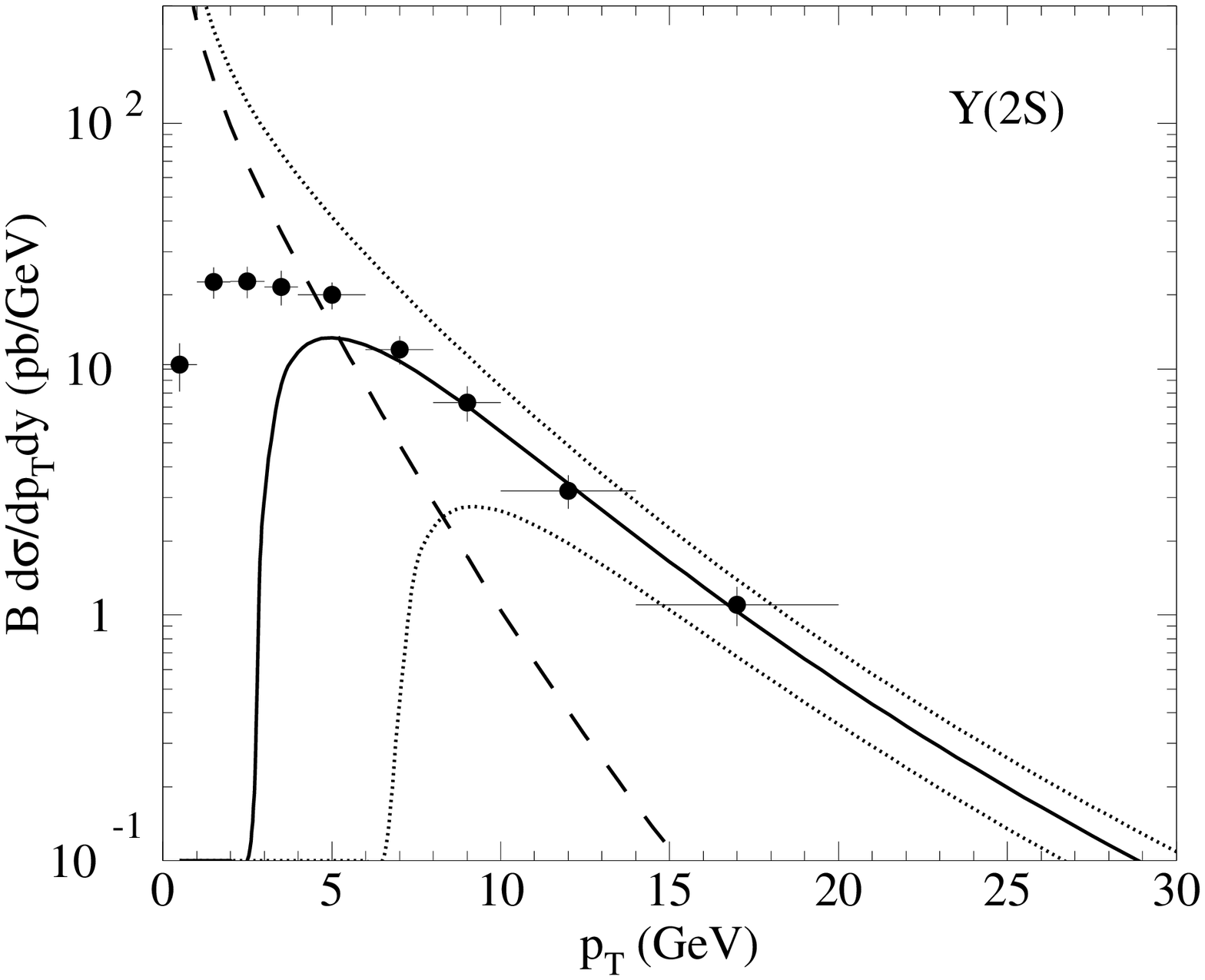}}
\caption[Inclusive cross section for $\Upsilon(2S)$ at Run I.]
{Inclusive cross section for $\Upsilon(2S)$ at $y = 0$ in Run I
	multiplied by its branching fraction $B$ into $\mu^+ \mu^-$
	as a function of $p_T$:  CDF data, 
	NRQCD fit, and color-singlet model prediction.}
\end{figure}

\begin{figure}[ht]
\centerline{\epsfxsize 11cm \epsffile{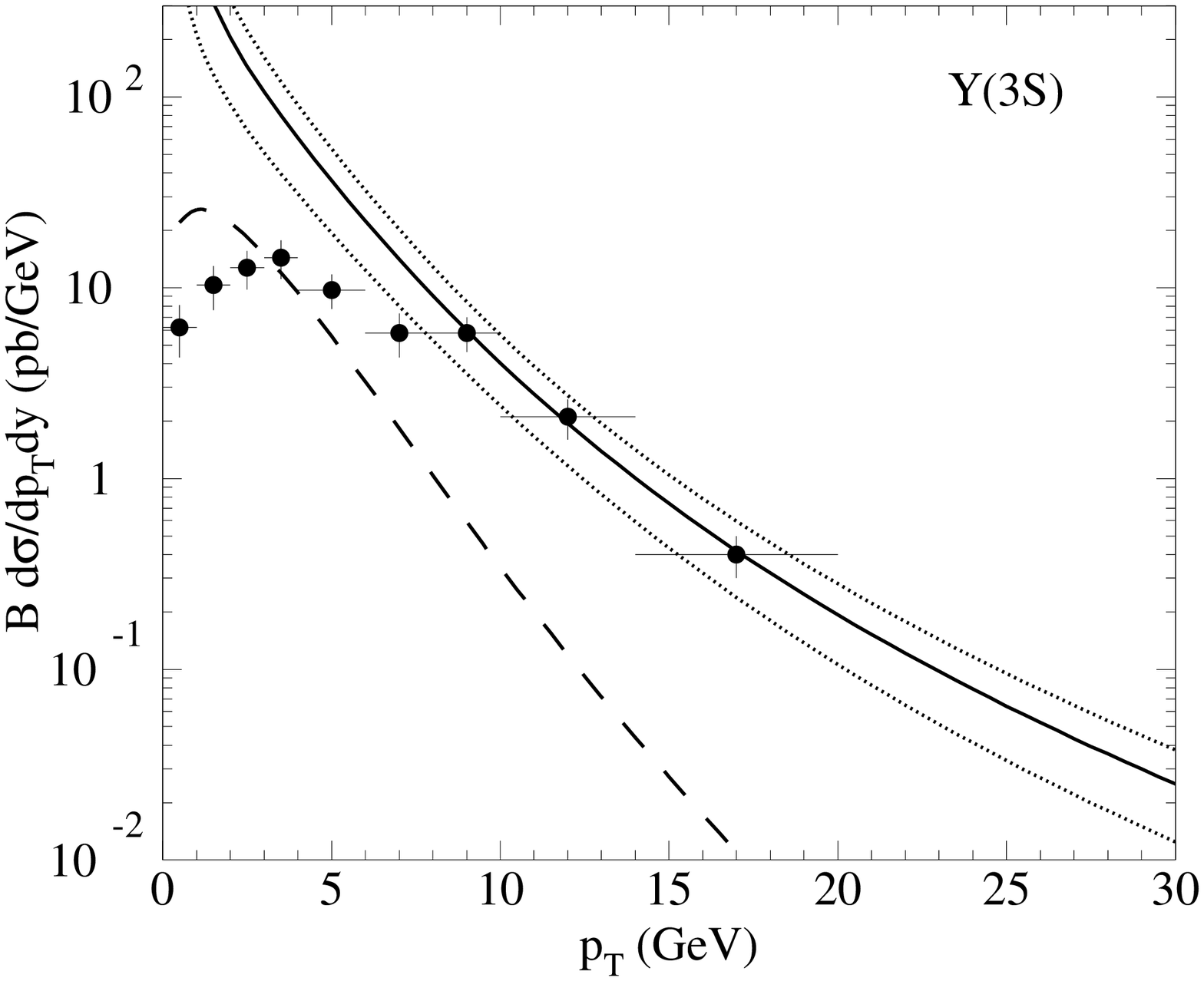}}
\caption[Inclusive cross section for $\Upsilon(3S)$ at $y = 0$ in Run I.]
{Inclusive cross section for $\Upsilon(3S)$ at Run I
	multiplied by its branching fraction $B$ into $\mu^+ \mu^-$
	as a function of $p_T$:  CDF data, 
	NRQCD fit, and color-singlet model prediction.}
\end{figure}

Our fits for $B d \sigma/d p_T d y$ at $y = 0$ for inclusive 
$\Upsilon(1S)$, $\Upsilon(2S)$, and $\Upsilon(3S)$ 
are compared to the CDF data in Figures 1, 2, and 3.
The error bands reflect the statistical uncertainties in the matrix 
elements.  The fits are reasonably good in the region
$p_T > 8$ GeV that we used for fitting.  
At low $p_T$, our fits
for $d \sigma/d p_T$ behave like $1/p_T$, because we have not 
implemented the effects of soft-gluon radiation 
on the shape of the $p_T$-distribution.
The fits therefore diverge from the data below $p_T = 8$ GeV.
For $\Upsilon(2S)$, the central curve becomes negative at small $p_T$
because our fit gives a negative central value 
for $\langle O_8(^1S_0) \rangle^{\Upsilon(2S)}_{\rm inc}$ 
or $\langle O_8(^3P_0) \rangle^{\Upsilon(2S)}_{\rm inc}$.
If we had fit the color-octet matrix elements using the data for 
$p_T > 4$ GeV, instead of only the data for $p_T > 8$ GeV,
the central values for $\langle O_8(^1S_0) \rangle_{\rm inc}$ 
or $\langle O_8(^3P_0) \rangle_{\rm inc}$ would also have been negative
for $\Upsilon(1S)$ and $\Upsilon(3S)$. 

In Figures 1, 2, and 3, the color-singlet model predictions from
order-$\alpha_s^3$ fusion processes are shown as dashed lines.  
At the largest values of $p_T$ shown, the predictions fall more than an 
order of magnitude below the data.
The color-singlet model prediction
for $\Upsilon(3S)$ indicates the shape of the $b \bar b_1(^3S_1)$ 
cross section.  The color-singlet model predictions for
$\Upsilon(1S)$ and $\Upsilon(2S)$ behave very differently 
at small $p_T$, because they receive contributions from decays of 
$\chi_{bJ}(nP)$.  The predictions diverge as $p_T \to 0$, 
because the cross sections $d \hat{\sigma}/dp_T$ for 
$b \bar b_1(^3P_{0,2})$ behave like $1/p_T$.
In order to obtain the correct threshold behavior in these channels,
it would be necessary to resum the effects of soft-gluon radiation.

\subsection{ Direct Matrix Elements for P-waves}

The color-octet matrix elements 
$\langle O_8(^3S_1) \rangle$ for the $\chi_b$'s
can be determined from CDF measurements of the fractions of 
$\Upsilon(1S)$'s that come 
from $\chi_b(1P)$'s and from $\chi_b(2P)$'s \cite{CDF-chib}.
The important feeddown decays for $\chi_{bJ}(1P)$ and $\chi_{bJ}(2P)$
proceed through the $\Upsilon(2S)$ and $\Upsilon(3S)$, respectively.
The fractions $F^{\Upsilon(1S)}_{\chi_b(nP)}$ of 
$\Upsilon(1S)$'s from $\chi_b(nP)$ therefore satisfy
\begin{eqnarray}
F^{\Upsilon(1S)}_{\chi_b(1P)} \; \sigma[\Upsilon(1S)]_{\rm inc} &=& 
\sum_J B_{\chi_{bJ}(1P) \to \Upsilon(1S)} \; \sigma[\chi_{bJ}(1P)]
\nonumber
\\
&& 
+ \left( \sum_J B_{\Upsilon(2S) \to \chi_{bJ}(1P)} 
		B_{\chi_{bJ}(1P) \to \Upsilon(1S)} \right) 
	\sigma[\Upsilon(2S)]_{\rm inc},
\label{f-chi1P}
\\
F^{\Upsilon(1S)}_{\chi_b(2P)} \; \sigma[\Upsilon(1S)]_{\rm inc} &=& 
\sum_J B_{\chi_{bJ}(2P) \to \Upsilon(1S)} \; \sigma[\chi_{bJ}(2P)] 
\nonumber
\\
&& 
+ \left( \sum_J B_{\Upsilon(3S) \to \chi_{bJ}(2P)} 
		B_{\chi_{bJ}(2P) \to \Upsilon(1S)} \right) 
	\sigma[\Upsilon(3S)]_{\rm inc}.
\label{f-chi2P}
\end{eqnarray}
Using the branching fractions from Table~\ref{tab:branch},
the coefficients of $\sigma[\Upsilon(2S)]_{\rm inc}$
and $\sigma[\Upsilon(3S)]_{\rm inc}$ in 
(\ref{f-chi1P}) and (\ref{f-chi2P}) are 
$(3.8 \pm 0.7)$\% and $(1.8 \pm 0.2)$\%, respectively.

The CDF result for the fractions of $\Upsilon(1S)$'s 
with $|y|<0.4$ and $p_T > 8$ GeV
that come from $\chi_b(1P)$'s and $\chi_b(2P)$'s are 
\begin{eqnarray}
F^{\Upsilon(1S)}_{\chi_b(1P)} &=& (27.1 \pm 8.1) \%,
\label{f-chi1Pexp}
\\
F^{\Upsilon(1S)}_{\chi_b(2P)} &=& (10.5 \pm 4.8)\%,
\label{f-chi2Pexp}
\end{eqnarray}
where we have added the statistical and systematic errors in quadrature.  
The inclusive $\Upsilon(nS)$ cross sections in (\ref{f-chi1P}) 
and (\ref{f-chi2P}) are the cross sections integrated over 
$|y| < 0.4$ and $p_T > 8$ GeV.  Using the CDF measurements 
of these cross sections, (\ref{f-chi1Pexp}) and (\ref{f-chi2Pexp})
reduce to the following constraints on the 
cross sections for $\chi_{bJ}(1P)$ and $\chi_{bJ}(2P)$:
\begin{eqnarray}
\sum_J B_{\chi_{bJ}(nP) \to \Upsilon(1S)} \; 
	\sigma[\chi_{bJ}(nP)] 
&=& (0.85 \pm 0.29) \; {\rm nb}, \qquad n=1,
\label{sig8-chi1}
\\
&=& (0.34 \pm 0.17) \; {\rm nb}, \qquad n=2.
\label{sig8-chi2}
\end{eqnarray}
The branching fractions and the associated errors are given in 
Table~\ref{tab:branch}.

The theoretical cross sections for $\sigma[\chi_{bJ}(nP)]$, $J=0,1,2$, are 
obtained by integrating (\ref{sig-chi}) over the appropriate region 
of $y$ and $p_T$.  The constraints (\ref{sig8-chi1}) and
(\ref{sig8-chi2}) are then linear equations for  
$\langle O^{\chi_{b0}(nP)}_8(^3S_1) \rangle$,
which give the values in Table~\ref{tab:CO-direct}.
The first error is obtained by setting $\mu_F = \mu_R = \mu_T$
and combining in quadrature the 
experimental error from (\ref{sig8-chi1}) or (\ref{sig8-chi2}), 
the error from the branching fractions in Table~\ref{tab:branch},
and the errors from the color-singlet matrix elements 
in Table~\ref{tab:CS-direct}.
The second upper and lower errors in Table~\ref{tab:CO-direct}
are the shifts in the matrix elements 
when $\mu_R$ and $\mu_F$ are changed from their central values 
by multiplicative factors of 2 and 1/2, respectively.

\begin {table}
\begin {center}
\begin {tabular}{r|c|c|c|c}
& \multicolumn{2}{c|}{CTEQ5L}
& \multicolumn{2}{c}{MRSTLO} \\
\hline
$\langle O^{\chi_{b0}(2P)}_8(^3S_1) \rangle$ 
& \multicolumn{2}{c|}{ $ 0.8 \pm 1.1^{+1.1}_{-0.8} $} 
& \multicolumn{2}{c}{ $ 1.2 \pm 1.3^{+1.1}_{-0.8} $} \\
\hline
$\langle O^{\chi_{b0}(1P)}_8(^3S_1) \rangle$ 
& \multicolumn{2}{c|}{ $ 1.5 \pm 1.1^{+1.3}_{-1.0} $} 
& \multicolumn{2}{c}{ $ 1.9 \pm 1.3^{+1.4}_{-1.0} $} \\
\hline
$\langle O^{\Upsilon(2S)}_8(^3S_1) \rangle$ 
& $ 16.4 \pm 5.7^{+7.1}_{-5.1} $&$ 15.6 \pm 5.2^{+6.9}_{-4.9}$ 
& $ 17.4 \pm 6.4^{+7.0}_{-5.1} $&$ 16.8 \pm 5.8^{+6.8}_{-5.0}$ 
\\
$\langle O^{\Upsilon(2S)}_8(^1S_0) \rangle$ 
& $ -10.8 \pm 9.7^{-3.4}_{+2.0} $ &   0   
& $ -9.5 \pm 11.1^{-2.8}_{+2.1} $ &   0   \\
${5 \over m_b^2} \langle O^{\Upsilon(2S)}_8(^3P_0) \rangle$ 
&  0  & $ -11.2 \pm 10.2^{-3.3}_{+2.4} $ 
&  0  & $ -9.7 \pm 11.6^{-2.9}_{+2.1} $ \\
\hline
$\langle O^{\Upsilon(1S)}_8(^3S_1) \rangle$ 
& $ 2.0 \pm 4.1^{-0.6}_{+0.5} $&$ 3.0 \pm 3.8^{+0.2}_{-0.1} $
& $ 0.4 \pm 4.7^{-1.0}_{+0.7} $&$ 1.8 \pm 4.4^{-0.2}_{+0.1} $
\\
$\langle O^{\Upsilon(1S)}_8(^1S_0) \rangle$ 
& $ 13.6 \pm 6.8^{+10.8}_{-7.5} $ &   0   
& $ 20.2 \pm 7.8^{+11.9}_{-8.5} $ &   0   \\
${5 \over m_b^2} \langle O^{\Upsilon(1S)}_8(^3P_0) \rangle$ 
&  0  & $ 13.9 \pm 7.1^{+11.4}_{-8.0} $ 
&  0  & $ 20.7 \pm 8.1^{+12.4}_{-8.8} $ \\
\end {tabular}
\end {center}
\caption{ \label {tab:CO-direct}
	Direct color-octet matrix elements for 
	$\chi_b$ and $\Upsilon$ states
	(in units of $10^{-2}$ GeV$^3$). }
\end {table}

\subsection{Direct Matrix Elements for S-waves}

The NRQCD matrix elements in Table~\ref{tab:CO-total} can be 
used to calculate the inclusive $\Upsilon(nS)$ cross sections.
To calculate the direct $\Upsilon(nS)$ cross sections, 
we must extract direct color-octet matrix elements 
for the $\Upsilon(nS)$ states from the inclusive color-octet 
matrix elements given in Table~\ref{tab:CO-total}.
The linear combinations of matrix elements determined by the
inclusive $\Upsilon(1S)$ cross sections are
\begin{eqnarray}
\langle O_8(^3S_1) \rangle^{\Upsilon(1S)}_{\rm inc} &=&
\langle O^{\Upsilon(1S)}_8(^3S_1) \rangle
+ 0.311 \; \langle O^{\Upsilon(2S)}_8(^3S_1) \rangle
+ 0.112 \; \langle O^{\Upsilon(3S)}_8(^3S_1) \rangle
\nonumber
\\
&&+ 2.15 \; \langle O^{\chi_{b0}(1P)}_8(^3S_1) \rangle
+ 1.08 \; \langle O^{\chi_{b0}(2P)}_8(^3S_1) \rangle,
\label{O83S-ups1S}
\\
\langle O_8(^1S_0) \rangle^{\Upsilon(1S)}_{\rm inc} &=&
\langle O^{\Upsilon(1S)}_8(^1S_0) \rangle
+ 0.311 \; \langle O^{\Upsilon(2S)}_8(^1S_0) \rangle
+ 0.112 \; \langle O^{\Upsilon(3S)}_8(^1S_0) \rangle,
\label{O81S-ups1S}
\\
\langle O_8(^3P_0) \rangle^{\Upsilon(1S)}_{\rm inc} &=&
\langle O^{\Upsilon(1S)}_8(^3P_0) \rangle
+ 0.311 \; \langle O^{\Upsilon(2S)}_8(^3P_0) \rangle
+ 0.112 \; \langle O^{\Upsilon(3S)}_8(^3P_0) \rangle.
\label{O83P-ups1S}
\end{eqnarray}
The errors on the branching fractions in 
(\ref{O83S-ups1S})--(\ref{O83P-ups1S}) have been suppressed, but they are
$(0.311 \pm 0.016)$, $(0.112 \pm 0.005)$, $(2.15 \pm 0.31)$,
and $(1.08 \pm 0.08)$.
The linear combinations determined by the
inclusive $\Upsilon(2S)$ cross sections are
\begin{eqnarray}
\langle O_8(^3S_1 \rangle^{\Upsilon(2S)}_{\rm inc} &=&
\langle O^{\Upsilon(2S)}_8(^3S_1) \rangle
+ 0.106 \; \langle O^{\Upsilon(3S)}_8(^3S_1) \rangle
+ 1.49 \; \langle O^{\chi_{b0}(2P)}_8(^3S_1) \rangle,
\label{O83S-ups2S}
\\
\langle O_8(^1S_0) \rangle^{\Upsilon(2S)}_{\rm inc} &=&
\langle O^{\Upsilon(2S)}_8(^1S_0) \rangle
+ 0.106 \; \langle O^{\Upsilon(3S)}_8(^1S_0) \rangle,
\label{O81S-ups2S}
\\
\langle O_8(^3P_0) \rangle^{\Upsilon(2S)}_{\rm inc} &=&
\langle O^{\Upsilon(2S)}_8(^3P_0) \rangle
+ 0.106 \; \langle O^{\Upsilon(3S)}_8(^3P_0) \rangle.
\label{O83P-ups2S}
\end{eqnarray}
The errors on the branching fractions in 
(\ref{O83S-ups2S})--(\ref{O83P-ups2S}) are
$(0.106 \pm 0.008)$ and $(1.49 \pm 0.17)$.
Using the color-octet matrix elements for inclusive $\Upsilon(nS)$
in Table~\ref{tab:CO-total} and the color-octet matrix 
elements for direct $\chi_b(nP)$ in Table~\ref{tab:CO-direct},
we obtain the color-octet matrix elements for direct 
$\Upsilon(nS)$ in Table~\ref{tab:CO-direct}.

Our analysis gives a negative value consistent with zero 
for the matrix elements 
$\langle O_8^{\Upsilon(2S)}(^1S_0) \rangle$ 
or $\langle O_8^{\Upsilon(2S)}(^3P_0) \rangle$.
Our values for $\langle O^{\chi_{b0}(2P)}_8(^3S_1) \rangle$, 
$\langle O^{\chi_{b0}(1P)}_8(^3S_1) \rangle$, 
and $\langle O^{\Upsilon(1S)}_8(^3S_1) \rangle$ 
are also consistent with zero given the statistical error.
The only direct color-octet matrix elements that differ from zero 
by two or more statistical error bars are 
$\langle O^{\Upsilon(2S)}_8(^3S_1) \rangle$ and 
$\langle O_8^{\Upsilon(1S)}(^1S_0) \rangle$ 
or $\langle O_8^{\Upsilon(1S)}(^3P_0) \rangle$.

We now compare our values for the matrix elements with those
obtained by Cho and Leibovich in their pioneering analysis of 
bottomonium production at the Tevatron.  
Their analysis was based on the CDF data sample from Run IA 
of the Tevatron \cite{CDF-upsIa}.  To reduce the 
errors associated with the 
shape of the $p_T$ distribution at small $p_T$, 
they used only the data from $p_T > 3.5$ GeV 
in their analysis.  The data was insufficient to determine 
all the matrix elements, so they estimated the matrix elements
$\langle O^{\Upsilon(nS)}_8(^3S_1) \rangle$ by applying scaling
relations to the corresponding matrix elements in the charmonium sector.
Their value for $\langle O^{\Upsilon(1S)}_8(^3S_1) \rangle$
is consistent with ours to within our large error bars, but their value
for $\langle O^{\Upsilon(2S)}_8(^3S_1) \rangle$ 
is smaller than ours by about a factor of 40. 
They used the CDF data to fit the matrix elements 
$\langle O_8^{\chi_b(nP)}(^3S_1) \rangle$ and the linear combinations 
$M_5^{\Upsilon(nS)} = \langle O_8^{\Upsilon(nS)}(^1S_0) \rangle 
	+ 5\langle O_8^{\Upsilon(nS)}(^3P_0) \rangle/m_b^2$.
Their values for $\langle O_8^{\chi_b(nP)}(^3S_1) \rangle$ 
are comparable to ours in magnitude, but they have much smaller error bars.
Their values for $M_5^{\Upsilon(nS)}$ differ from zero by only about 
one error bar.  In our analysis, we included the matrix elements
$\langle O^{\Upsilon(nS)}_8(^3S_1) \rangle$ in the list of those to
be fit to the CDF data. The much higher quality of the CDF data 
from Run IB of the Tevatron \cite{CDF-upsIb} allowed us to carry out a 
reasonable fit using the data restricted to $p_T > 8$ GeV.

\section{Inclusive Cross Sections for Spin-triplet States}

Having determined the most important matrix elements
for the production of the spin-triplet bottomonium states,
we can use them to calculate the cross sections for these states
in other high energy processes. In particular, 
we can calculate their cross sections in Run II of the Tevatron
in which the center-of-mass energy will be increased from
1.8 TeV to 2.0 TeV.
In order to cancel the large theoretical errors, 
such as those from the uncertainties in the matrix elements 
and from the choice of scale,
we normalize the cross sections to that for inclusive $\Upsilon(1S)$ 
at $\sqrt{s}=1.8$ TeV.  For each bottomonium state $H$, we define the ratio
\begin{eqnarray}
R^H(\sqrt{s})
&=&  {\sigma[H; \; \sqrt{s}] \over
	\sigma[{\rm inclusive} \; \Upsilon(1S); \; \sqrt{s}=1.8 \; {\rm TeV}]},
\label{R-H}
\end{eqnarray}
where the cross sections are integrated over $p_T> 8$ GeV 
and over $|y|< 0.4$.

To calculate the inclusive cross section for $\Upsilon (nS)$,
we simply use the inclusive color-singlet matrix elements from 
Table~\ref{tab:CS-total} and the inclusive color-octet matrix elements from 
Table~\ref{tab:CO-total}.
To calculate the inclusive cross section for  $\chi_{bJ}(nP)$,
we must first compute the direct cross sections for 
$\chi_{bJ}(nP)$ and the higher bottomonium states using the 
direct color-singlet matrix elements from 
Table~\ref{tab:CS-direct} and the direct color-octet matrix elements from 
Table~\ref{tab:CO-direct}, and then combine them using the inclusive branching 
fractions in Table~\ref{tab:branch}.
The resulting ratios $R^H$ shown in Table~\ref{tab:Rincl}
are the averages of the 4 values obtained by using either the 
CTEQ5L or MRST98LO parton distributions and either setting
$\langle O_8(^1S_0) \rangle=0$ 
or $\langle O_8(^3P_0) \rangle=0$.
The error bars come from combining in quadrature the statistical errors 
in the matrix elements from the Tables,
the error from varying $\mu$ by a factor of two from its central value, 
the difference between using the CTEQ5L and MRST98LO parton distributions,
the difference between setting $\langle O_8(^1S_0) \rangle=0$ 
and $\langle O_8(^3P_0) \rangle=0$, 
and the error from varying $m_b$.
The error bars in the numerator and denominator of (\ref{R-H})
are both large, but they are highly correlated and tend to cancel 
in the ratio.  
The largest contributions to the error bars are the statistical errors 
on the matrix elements, with the exception of $\Upsilon(1S)$,
for which the largest contribution comes from varying the scale. 
For $\Upsilon(2S)$ and $\Upsilon(3S)$, 
the results in Table~\ref{tab:Rincl} for $\sqrt{s} = 1.8$ TeV
are consistent with the actual CDF measurements, which give
$0.61 \pm 0.12$ for $\Upsilon(2S)$ and $0.29 \pm 0.12$ for $\Upsilon(3S)$.
When the center-of-mass energy is increased from 1.8 TeV to 2.0 TeV,
all the cross sections increase by about 16\%.
The increase depends on $p_T$, changing from about 15\% at $p_T = 8$ TeV
to  about 19\% at $p_T = 20$ TeV.

\begin {table}[t]
\begin {center}
\begin {tabular}{l|cc|cc}
$H$             & $R^H$(1.8 TeV) && $R^H$(2.0 TeV) \\
\hline
$\Upsilon(3S)$  & $0.31 \pm 0.14$ && $0.36 \pm 0.16$ \\
\hline
$\chi_{b2}(2P)$ & $0.44 \pm 0.26$ && $0.52 \pm 0.30$ \\
$\chi_{b1}(2P)$ & $0.34 \pm 0.16$ && $0.39 \pm 0.19$ \\
$\chi_{b0}(2P)$ & $0.20 \pm 0.07$ && $0.24 \pm 0.08$ \\
\hline
$\Upsilon(2S)$  & $0.65 \pm 0.35$ && $0.76 \pm 0.41$ \\
\hline
$\chi_{b2}(1P)$ & $0.57 \pm 0.26$ && $0.66 \pm 0.31$ \\
$\chi_{b1}(1P)$ & $0.41 \pm 0.17$ && $0.48 \pm 0.19$ \\
$\chi_{b0}(1P)$ & $0.23 \pm 0.08$ && $0.26 \pm 0.09$ \\
\hline
$\Upsilon(1S)$  &        1        && $1.16 \pm 0.01$ 
\end {tabular}
\end {center}
\caption{ \label {tab:Rincl}
	Ratios of the inclusive cross sections for the spin-triplet 
	bottomonium states $H$ at the Tevatron 
	with $\sqrt{s}=$ 1.8 TeV and 2.0 TeV to the 
	inclusive cross section for $\Upsilon(1S)$
	with $\sqrt{s}=$ 1.8 TeV. }
\end {table}

\section{Direct Cross Sections for Spin-singlet States}

Having determined the most important matrix elements
for the production of the spin-triplet bottomonium states,
we can also use them 
to calculate the production rate of the spin-singlet 
states $\eta_b(nS)$ and $h_b(nP)$.
The matrix elements for these states are related to those of the 
corresponding spin-triplet states by 
the approximate spin symmetry of NRQCD.  
Spin symmetry relates the matrix elements for
$\eta_b(nS)$ to those for $\Upsilon(nS)$:
\begin{eqnarray}
\langle O^{\eta_b(nS)}_1(^1S_0) \rangle
&=&  {1 \over 3} \langle O^{\Upsilon(nS)}_1(^3S_1) \rangle,
\label{eta-ss1}
\\
\langle O^{\eta_b(nS)}_8(^1P_1) \rangle
&=& {1 \over 3} \langle O^{\Upsilon(nS)}_8(^3P_0) \rangle,
\label{eta-ss2}
\\
\langle O^{\eta_b(nS)}_8(^1S_0) \rangle
&=&  {1 \over 3} \langle O^{\Upsilon(nS)}_8(^3S_1) \rangle,
\label{eta-ss3}
\\
\langle O^{\eta_b(nS)}_8(^3S_1) \rangle
&=&  {1 \over 3} \langle O^{\Upsilon(nS)}_8(^1S_0) \rangle.
\label{eta-ss4}
\end{eqnarray}
The direct matrix elements for $\eta_b(nS)$ can therefore be 
read off from those for $\Upsilon(nS)$ in Tables~\ref{tab:CS-direct}
and \ref{tab:CO-direct}.
The spin-symmetry relations have been used 
to calculate the cross sections for producing the $\eta_c$ at the 
Tevatron \cite{M-P-S}, in photoproduction and electroproduction, 
\cite{H-Y-C}, and at Hera-B \cite{Qiao-Yuan}.
In Refs. \cite{M-P-S} and \cite{H-Y-C}, the factor of ${1 \over 3}$
which comes from the number of spin states
was omitted from the spin-symmetry relations (\ref{eta-ss1})-(\ref{eta-ss4}).
Thus their estimates of the cross sections may be too large by a factor of 3.
Spin symmetry relates the matrix elements for $h_b(nP)$ 
to those for $\chi_{b0}(nP)$:
\begin{eqnarray}
\langle O^{h_b(nP)}_1(^1P_1) \rangle
&=&  3  \langle O^{\chi_{b0}(nP)}_1(^3P_0) \rangle,
\label{h-ss1}
\\
\langle O^{h_b(nP}_8(^1S_0) \rangle
&=&  3 \langle O^{\chi_{b0}(nP)}_8(^3S_1) \rangle.
\label{h-ss2}
\end{eqnarray}
The direct matrix elements for $h_b(nP)$ can therefore be 
read off from those for $\chi_b(nP)$ in Tables~\ref{tab:CS-direct}
and \ref{tab:CO-direct}.
These relations were first used by Fleming and Mehen 
to calculate the cross section for 
the photoproduction of $h_c$ \cite{F-M}.
They have also been used to calculate the cross sections for $h_c$
at Hera-B \cite{Qiao-Yuan}.

To calculate the cross sections for the direct production of the spin-singlet 
states at the Tevatron, we need the appropriate parton differential cross 
sections $d\hat \sigma$.  Explicit expressions for most of those that are 
needed are given in Ref. \cite{Cho-Leibovich} and in Ref. \cite{B-K-V}.
The exception is the cross sections for producing $b \bar b_8(^1P_1)$, 
which contributes to $\eta_b$ production.
The cross sections for the production of $b \bar b_8(^1P_1)$ in
$q \bar q$, $gq$, and $gg$ scattering are 
\begin{eqnarray}
\frac{d\sigma}{d\hat{t}}\left(q\bar{q}\to \, ^1P_1^{(8)}g\right) &=&
   \frac{\langle O_8(^1P_1) \rangle}{16\pi\hat{s}^2}
   \frac{4(4\pi\alpha_s)^3}{9M^3}
   \frac{\hat{t}^2+\hat{u}^2}{\hat{s}(\hat{s}-M^2)^2},
\\
\frac{d\sigma}{d\hat{t}}\left(gq\to \, ^1P_1^{(8)}q\right) &=&
   \frac{\langle O_8(^1P_1)\rangle}{16\pi\hat{s}^2}
   \frac{(4\pi\alpha_s)^3}{6M^3}
   \frac{\hat{s}^2+\hat{u}^2}{(-\hat{t})(M^2-\hat{t})^2},
\\
\frac{d\sigma}{d\hat{t}}\left(gg\to ^1P_1^{(8)}g\right) &=&
   \frac{\langle O_8(^1P_1) \rangle}{16\pi\hat{s}^2}
   \frac{(4\pi\alpha_s)^3}{36M^3}
   \frac{1}{z^2\hat{s}(\hat{s}-M^2)^3(\hat{s} M^2+z^2)^3}
\nonumber\\
&&\left\{ 
 27 \hat{s} z^2(\hat{s}^8-4\hat{s}^6z^2+\hat{s}^4z^4+\hat{s}^2z^6+z^8)
\right.
\nonumber\\
&&\,
 +M^2 (27\hat{s}^{10} - 243\hat{s}^8 z^2 + 697\hat{s}^6 z^4
   - 665 \hat{s}^4 z^6 + 346 \hat{s}^2 z^8 - 27 z^{10})
\nonumber\\
&&\,
 -M^4\hat{s} (135\hat{s}^8 - 702\hat{s}^6 z^2 + 1340 \hat{s}^4 z^4 
   - 1087 \hat{s}^2 z^6 + 135 z^8)\nonumber\\
&&\,
 +M^6 (324\hat{s}^8 -1134\hat{s}^6 z^2 + 1557 \hat{s}^4 z^4 
   - 698 \hat{s}^2 z^6 + 54 z^8)
\nonumber\\
&&\,
 -M^8 \hat{s} (486 \hat{s}^6 - 1091\hat{s}^4 z^2 + 882 \hat{s}^2 z^4
   - 92 z^6)
\nonumber\\
&&\,
 +M^{10} (486 \hat{s}^6 - 616 \hat{s}^4 z^2 + 374 \hat{s}^2 z^4 
   - 27 z^6)\nonumber\\
&&\,
 -M^{12}\hat{s} (324 \hat{s}^4 - 211 \hat{s}^2 z^2 + 38 z^4)
\nonumber\\
&&\,
\left. + M^{14} \hat{s}^2(135 \hat{s}^2 - 38 z^2) - 27 M^{16}\hat{s}^3\right\}.
\end{eqnarray}
%
where $z^2 = \hat{t}\hat{u}$. 

We proceed to calculate the cross sections 
for the direct production of the spin-singlet states at the Tevatron
at center-of-mass energies $\sqrt{s} = 1.8$ TeV and 2.0 TeV.
To minimize the effect of the highly correlated errors,
we calculate the ratio (\ref{R-H}) of the direct cross section
integrated over $p_T> 8$ GeV and over $|y|< 0.4$ to the 
corresponding inclusive cross section for $\Upsilon (1S)$ at 1.8 TeV.
The resulting predictions are shown in Table~\ref{tab:Rincl}.
The cross sections for the $h_b(nP)$ states are small 
compared to those for $\Upsilon(1S)$ and they have large error bars.
The cross sections for $\eta_b(nS)$ are predicted to be
several times larger than those for $\Upsilon(nS)$ and they have
reasonably small error bars.
These predictions should be fairly reliable, 
because the cross sections are for the same kinematical region
as the data used to extract the matrix elements.
When the center-of-mass energy is increased from 1.8 TeV to 2 TeV, 
all the cross sections increase by about 16\%.

\begin {table}
\begin {center}
\begin {tabular}{l|cc|cc}
$H$             & $R^H$(1.8 TeV) && $R^H$(2.0 TeV) \\
\hline
$\eta_b(3S)$    & $1.72 \pm 0.52$ && $2.00 \pm 0.61$ \\
\hline
$h_b(2P)$       & $0.07 \pm 0.07$ && $0.08 \pm 0.09$ \\
\hline
$\eta_b(2S)$    & $1.77 \pm 0.50$ && $2.06 \pm 0.59$ \\
\hline
$h_b(1P)$       & $0.11 \pm 0.08$ && $0.13 \pm 0.09$ \\
\hline
$\eta_b(1S)$    & $4.31 \pm 0.98$ && $5.02 \pm 1.14$ \\
\end {tabular}
\end {center}
\caption{ \label {tab:Rdirect}
	Ratios of the direct cross sections for the spin-singlet 
	bottomonium states $H$ at the Tevatron 
	with $\sqrt{s}=$ 1.8 TeV and 2.0 TeV to the 
	inclusive cross section for $\Upsilon(1S)$
	with $\sqrt{s}=$ 1.8 TeV. }
\end {table}

We can make a rough estimate of the cross sections integrated 
over all $p_T$ by assuming that the spin-singlet cross sections 
have the same shape at small $p_T$  as the $\Upsilon(1S)$ cross section.
The measured inclusive $\Upsilon(1S)$ cross section at central rapidity 
integrated over all $p_T$ up to 20 GeV satisfies
$B d\sigma/dy = 690 \pm 25$ pb, where $B\approx 2.5\%$ 
is the branching fraction of $\Upsilon(1S)$ into $\mu^+ \mu^-$.
The cross section integrated only over $p_T > 8$ GeV satisfies
$B d\sigma/dy = 106 \pm 7$ pb.  The ratio of these cross sections 
is $6.5 \pm 0.5$. 
Multiplying the inclusive $\Upsilon(1S)$ cross section 
$d\sigma/dy = 28$ nb by the factor of 6.5 to take into account 
the small $p_T$ region and by the ratio 4.3 from Table~\ref{tab:Rincl},
we find that the cross section for $\eta_b(1S)$ integrated over all $p_T$
should be approximately $d\sigma/dy = 800$ nb.

The cross section for $\eta_b(1S)$ indicates that this state 
must have been produced in abundance in Run I of the Tevatron.
However the $\eta_b(1S)$ can be observed only if it has a large enough 
branching fraction into a decay mode that can be triggered upon.
One possibility is the double-$J/\psi$ decay 
$\eta_b(1S) \to J/\psi + J/\psi$, followed by the decays
$J/\psi \to \mu^+ \mu^-$.
The decay $\eta_b(1S) \to J/\psi + J/\psi$ has essentially the same 
kinematics as the decay $\eta_c \to \phi \phi$, 
except that all masses are scaled up by a factor of 3.
Thus the branching fraction for $\eta_b(1S) \to J/\psi + J/\psi$ 
could be as large as that for $\eta_c \to \phi \phi$,
which is approximately $7 \times 10^{-3}$.
We can obtain a lower bound on the branching fraction by using
the fact that in the limit $m_b \to \infty$ with $m_c$ fixed, 
the branching fraction for $\eta_b(1S) \to J/\psi + J/\psi$ 
scales like $1/m_b^4$ \cite{B-L}.
If $m_c$ and $m_b$ were both in this scaling region, 
then the branching fraction into light $J^{PC} = 1^{--}$ mesons 
would be smaller for $\eta_b(1S)$ than for $\eta_c$ by a factor of
$(m_c/m_b)^4$, which is about $10^{-2}$.  Since we are not deep 
into this scaling region, the suppression should be smaller than this.
Thus the branching fraction for $\eta_b(1S) \to J/\psi + J/\psi$ 
should be in the range between $7 \times 10^{-5}$ and $7 \times 10^{-3}$.
Multiplying by the branching fractions of 6\% for each of the decays
$J/\psi \to \mu^+ \mu^-$, our estimate for the branching fraction for 
$\eta_b(1S) \to J/\psi + J/\psi \to \mu^+ \mu^- + \mu^+ \mu^-$
is $B \approx 2.5 \times 10^{-6 \pm 1}$.  
The cross section for producing
this particular decay mode of $\eta_b(1S)$ is therefore 
$B d\sigma/dy \approx 2$ pb, give or take a factor of 10.  
Multiplying by the rapidity interval 0.8 and by the integrated luminosity 
of about 100 pb$^{-1}$ in Run I of the Tevatron, 
we obtain between 16 and 1600 produced events.

We must also take into account the acceptances and efficiencies 
for observing the decays $J/\psi \to \mu^+ \mu^-$.
These can be estimated using the CDF data on the production  of
prompt $J/\psi$ in Run IA of the Tevatron \cite{CDF-psi}.
Based on the observation of about 22,000 $J/\psi \to \mu^+ \mu^-$ candidates 
with $p_T > 5$ GeV and pseudorapidity $|\eta| < 0.6$ in a
15 pb$^{-1}$ data sample, they measured the cross section
in that region of $p_T$ and $\eta$ to be $B \sigma \approx 17$ nb.
We infer that the product of the acceptance and the
efficiency is roughly $\epsilon \approx 0.09$.
Multiplying the number of events that are produced by $\epsilon^2$, 
we get between 0.13 and 13 observed events.
Thus this cross section may be large enough to be observed 
in Run I of the Tevatron.  In Run II,
the integrated luminosity will be larger by a factor of 20 and there
will be significant improvements in the acceptances and 
efficiencies for observing muons in both the CDF and D0 detectors.
Thus the $\eta_b(1S)$ should certainly be observed in Run II
through the decay $\eta_b \to J/\psi + J/\psi$.

\section{Discussion}

We have carried out an updated NRQCD analysis of the
CDF data on the production of spin-triplet bottomonium states
from Run I of the Tevatron.
In spite of using only the data from $p_T > 8$ GeV,
we were able to extract all the relevant 
color-octet matrix elements directly from the data.
Only one of the 8 color-octet matrix elements comes out 
with a negative central value, but several others are also 
consistent with zero to within errors.
In our analysis, we distinguished between the inclusive 
color-octet matrix elements that can be used to compute inclusive 
$\Upsilon(nS)$ cross sections and the direct
color-octet matrix elements required to compute direct 
$\Upsilon(nS)$ cross sections and, by spin symmetry, direct 
$\eta_b(nS)$ cross sections.

The most serious deficiency in our analysis was our failure 
to take into account the effects of soft-gluon radiation that 
are needed to give a smooth $p_T$ distribution at small $p_T$.
This forced us to use only the small fraction of the data from 
$p_T > 8$ GeV to fit the color-octet matrix elements, which
led to large errors in these matrix elements.
If these matrix elements are used to predict bottomonium 
cross sections in other high energy processes, the 
predictions will probably have large error bars.
Our theoretical cross sections also diverge from the CDF data 
below $p_T = 8$ GeV, which gives us another reason to be
cautious in applying our matrix elements
to other high energy processes.

An analysis that deals properly with the small $p_T$ region 
could take full advantage of the CDF data and therefore 
determine the color-octet matrix elements more accurately.
Such an analysis requires a prescription for combining the 
leading-order cross sections for $i j \to b \bar b + k$
with the next-to-leading order cross sections for 
$i j \to b \bar b$ recently calculated by Petrelli et al. 
\cite{PCGMM} to get a smooth $p_T$ distribution near $p_T =0$.  
The matrix elements extracted from such an analysis should 
give reliable predictions for observables 
at the Tevatron that are dominated by low $p_T$.
They should also give reliable predictions for
bottomonium production in other high energy processes.

Our analysis should give reliable predictions for the cross sections 
of the spin-singlet states $\eta_b(nS)$ and $h_b(nP)$
at the Tevatron.  We find that the direct cross section for 
$\eta_b(1S)$ at $p_T > 8$ GeV should be greater than the inclusive 
cross section for $\Upsilon(1S)$ by a factor of about 4.3.
If we assume that the $p_T$ distributions have the same shapes 
at smaller $p_T$, we can estimate the cross section for $\eta_b(1S)$
integrated over all $p_T$.  The resulting cross section is large 
enough that it should be possible to discover the $\eta_b(1S)$ 
at the Tevatron if there is a decay mode with a large enough 
branching fraction that can be used as a trigger.  
We argued that the decay $\eta_b \to J/\psi + J/\psi$ 
should allow the discovery of the $\eta_b(1S)$, 
if not in the data from Run I, then certainly in Run II of the Tevatron.

\acknowledgements

This work was supported in part by the 
U.S. Department of Energy under Grants DE-FG02-91-ER40690
and DOE-ER-40682-143 and by the Natural Sciences and
Engineering Research Council of Canada.
We thank F. Maltoni for valuable comments.


\end{document}